# COMPUTE THE EDGE P-LAPLACIAN CENTRALITY FOR AIR TRAFFIC NETWORK

Tran Hoang Loc[1*], Tran Nguyen Bao[1], Nguyen Luong Anh Tuan[1]

[1]Vietnam Aviation Academy, Vietnam

*Corresponding Author/Email: locth@vaa.edu.vn

**ABSTRACT**

The problem that we would like to solve in this paper is to compute the edge p-Laplacian centrality for the air traffic network. In this problem, instead of computing the edge p-Laplacian centrality directly which is the very hard problem, we convert the air traffic network to the line graph. Finally, we will compute the node p-Laplacian centrality of the line graph which is equivalent to the edge p-Laplacian of the air traffic network. In this paper, the novel un-normalized graph (p-) Laplacian based ranking method will be developed based on the un-normalized graph p-Laplacian operator definitions such as the curvature operator of graph (i.e. the un-normalized graph 1-Laplacian operator) and will be used to compute the node p-Laplacian centrality of the line graph. The results from the experiments show that the un-normalized graph p-Laplacian ranking methods can be implemented successfully.

## 1. Introduction

Node centrality in air traffic networks is a valuable tool for understanding and optimizing the complex relationships between airports. Here are the key motivations for using node centrality:

**a. Identifying Critical Hub Airports:**
- **Degree Centrality:** Identifies airports with the highest number of direct connections, indicating their importance as hubs [1].
- **Betweenness Centrality:** Pinpoints airports that lie on the shortest paths between many other airports, highlighting their role in connecting different parts of the network [2].
- **Closeness Centrality:** Measures how quickly an airport can reach all other airports, indicating its accessibility and potential influence [3].

**b. Assessing Network Resilience and Vulnerability:**
- **Identifying Bottlenecks:** High-centrality airports are potential bottlenecks. Disruptions at these airports can have cascading effects on the entire network.
- **Assessing the Impact of Disruptions:** Simulating the removal of high-centrality airports can help assess the network's airports can help assess the network's vulnerabilities.

**c. Optimizing Network Efficiency:**
- **Routing and Scheduling:** Identifying high-centrality airports can help optimize flight routes and schedules, reducing travel times and costs.
- **Capacity Planning:** Understanding the importance of different airports can inform decisions about capacity expansion and infrastructure investments.

While node centrality focuses on the importance of individual airports, edge centrality provides insights into the significance of specific flight routes within the air traffic network. Here are the key motivations for using edge centrality in addition to node centrality:

**a. Identifying Critical Flight Routes:**
- **Edge Betweenness Centrality:** Identifies the most frequently used flight routes,

highlighting their importance for network connectivity and efficiency [4].
- **Edge Strength Centrality:** Measures the capacity or volume of traffic on a particular route, indicating its significance for passenger and cargo flow [5].

**b. Assessing Network Resilience and Vulnerability:**
- **Identifying Weak Links:** Low-centrality edges represent potential vulnerabilities in the network, as disruptions on these routes can significantly impact connectivity.
- **Assessing the Impact of Disruptions:** Simulating the removal of high-centrality edges can help identify critical routes and their potential impact on network performance.

**c. Optimizing Network Efficiency:**
- **Routing and Scheduling:** Identifying high-centrality edges can help optimize flight schedules and allocate resources more effectively.
- **Capacity Planning:** Understanding the importance of different routes can inform decisions about capacity expansion and infrastructure investments.

By combining node and edge centrality analysis, aviation stakeholders can gain a more comprehensive understanding of the air traffic network, especially in the research field air flight delay prediction. Inspired from the work including the node centralities (for example, PageRank centrality of the air traffic network) as the additional features in the regression algorithm (i.e., to solve the air flight delay prediction problem) [6, 7, 8, 9, 10, 11, 12, 13, 14, 15, 16, 17, 18, 19, 20, 21, 22, 23, 24, 25, 26], in this paper, we propose the novel method computing the p-Laplacian centrality for the flight (i.e., the edge of the air traffic network). This novel centrality, in the end, can be employed as the novel feature for the regression algorithm. In detail, in this paper, the un-normalized graph (p-) Laplacian based ranking method will be developed (for the edge of the air traffic network) based on the un-normalized graph p-Laplacian operator definitions such as the curvature operator of graph (i.e. the un-normalized graph 1-Laplacian operator) [27, 28].

We will organize the paper as follows: Section 2 will introduce the preliminary notations and definitions used in this paper. Section 3 will introduce the definition of the gradient and divergence operators of graphs. Section 4 will introduce the definition of Laplace operator of graphs and its properties. Section 5 will introduce the definition of the curvature operator of graphs and its properties. Section 6 will introduce the definition of the p-Laplace operator of graphs and its properties. Section 7 will show how to derive the algorithm of the un-normalized graph p-Laplacian based ranking method from regularization framework. In section 8, we will show the experimental results of the PageRank algorithm and the un-normalized graph p-Laplacian based ranking algorithms. Section 9 will conclude this paper, and the future direction of research will be discussed.

## 2. Preliminary notations and definitions

Suppose that you are given a graph $G' = (V', E', W')$ which is the air traffic network in this paper.

Instead of computing the PageRank centrality or the p-Laplacian centrality for the edge of the graph $G'$, which is the very hard problem, we convert the graph $G'$ to the line graph like the following:
- Each vertex of the line graph represents an edge of $G'$
- Two vertices in the line graph are connected by an edge if and only if their corresponding edges in $G'$ share a common vertex.

Let the line graph be $G$.

The graph $G = (V, E, W)$ where $V$ is a set of vertices with $|V| = n$, $E \subseteq V * V$ is a set of edges and $W$ is a $n * n$ similarity matrix with elements $w_{ij} > 0$ $(1 \leq i, j \leq n)$.

Also, please note that $w_{ij} = w_{ji}$.

The degree function $d: V \to R^+$ is
$$d_i = \sum_{j \sim i} w_{ij}, (1)$$
where $j \sim i$ is the set of vertices adjacent with $i$.

Define $D = diag(d_1, d_2, \ldots, d_n)$.

The inner product on the function space $R^V$ is
$$< f, g >_V = \sum_{i \in V} f_i g_i \ (2)$$

Also define an inner product on the space of functions $R^E$ on the edges
$$< F, G >_E = \sum_{(i,j) \in E} F_{ij} G_{ij} \ (3)$$

Here let $H(V) = (R^V, <.,.>_V)$ and $H(E) = (R^E, <.,.>_E)$ be the Hilbert space real-valued functions defined on the vertices of the graph $G$ and the Hilbert space of real-valued functions defined in the edges of $G$ respectively.

Thus, instead of computing the p-Laplacian centrality for edge of the graph $G'$, we can compute the p-Laplacian centrality for the node of the line graph $G$ of $G'$. This is the novelty of our work. Please note that the PageRank centrality of the line graph $G$ will be served as the baseline method for this paper. The following sections present how to compute the p-Laplacian centrality of the node of the graph.

## 3. Gradient and Divergence Operators

We define the gradient operator $d: H(V) \to H(E)$ to be
$$(df)_{ij} = \sqrt{w_{ij}}(f_j - f_i), (4)$$

where $f: V \to R$ be a function of $H(V)$.

We define the divergence operator $div: H(E) \to H(V)$ to be

$$< df, F >_{H(E)} = < f, -divF >_{H(V)}, (5)$$

where $f \in H(V), F \in H(E)$

Next, we need to prove that

$$(divF)_j = \sum_{i \sim j} \sqrt{w_{ij}} (F_{ji} - F_{ij})$$

Proof:

$$< df, F > = \sum_{(i,j) \in E} df_{ij} F_{ij}$$
$$= \sum_{(i,j) \in E} \sqrt{w_{ij}} (f_j - f_i) F_{ij}$$
$$= \sum_{(i,j) \in E} \sqrt{w_{ij}} f_j F_{ij} - \sum_{(i,j) \in E} \sqrt{w_{ij}} f_i F_{ij}$$
$$= \sum_{k \in V} \sum_{i \sim k} \sqrt{w_{ik}} f_k F_{ik} - \sum_{k \in V} \sum_{j \sim k} \sqrt{w_{kj}} f_k F_{kj}$$
$$= \sum_{k \in V} f_k (\sum_{i \sim k} \sqrt{w_{ik}} F_{ik} - \sum_{i \sim k} \sqrt{w_{ki}} F_{ki})$$
$$= \sum_{k \in V} f_k \sum_{i \sim k} \sqrt{w_{ik}} (F_{ik} - F_{ki})$$

Thus, we have

$$(divF)_j = \sum_{i \sim j} \sqrt{w_{ij}} (F_{ji} - F_{ij}) \quad (6)$$

## 4. Laplace operator

We define the Laplace operator $\Delta: H(V) \to H(V)$ to be

$$\Delta f = -\frac{1}{2} div(df) \quad (7)$$

Next, we compute

$$(\Delta f)_j = \frac{1}{2} \sum_{i \sim j} \sqrt{w_{ij}} ((df)_{ij} - (df)_{ji})$$
$$= \frac{1}{2} \sum_{i \sim j} \sqrt{w_{ij}} (\sqrt{w_{ij}}(f_j - f_i) - \sqrt{w_{ij}}(f_i - f_j))$$
$$= \sum_{i \sim j} w_{ij} (f_j - f_i)$$
$$= \sum_{i \sim j} w_{ij} f_j - \sum_{i \sim j} w_{ij} f_i$$
$$= d_j f_j - \sum_{i \sim j} w_{ij} f_i$$

Thus, we have

$$(\Delta f)_j = d_j f_j - \sum_{i \sim j} w_{ij} f_i \quad (8)$$

The graph Laplacian is a linear operator. Furthermore, the graph Laplacian is self-adjoint and positive semi-definite.

Let $S_2(f) = < \Delta f, f >$, we have the following **theorem 1**

$$D_f S_2 = 2\Delta f \quad (9)$$

The proof of the above theorem can be found from [27, 28]

## 5. Curvature operator

We define the curvature operator $\kappa: H(V) \to H(V)$ to be

$$\kappa f = -\frac{1}{2} div(\frac{df}{||df||}) \quad (10)$$

Next, we compute

$$(\kappa f)_j = \frac{1}{2} \sum_{i \sim j} \sqrt{w_{ij}} ((\frac{df}{||df||})_{ij} - (\frac{df}{||df||})_{ji})$$
$$= \frac{1}{2} \sum_{i \sim j} \sqrt{w_{ij}} (\frac{1}{||d_i f||} \sqrt{w_{ij}}(f_j - f_i) - \frac{1}{||d_j f||} \sqrt{w_{ij}}(f_i - f_j))$$
$$= \frac{1}{2} \sum_{i \sim j} w_{ij} (\frac{1}{||d_i f||} + \frac{1}{||d_j f||})(f_j - f_i)$$

Thus, we have

$$(\kappa f)_j = \frac{1}{2} \sum_{i \sim j} w_{ij} (\frac{1}{||d_i f||} + \frac{1}{||d_j f||})(f_j - f_i) \quad (11)$$

From the above formula, we have

$$d_i f = ((df)_{ij} : j \sim i)^T \quad (12)$$

The local variation of $f$ at $i$ is defined to be

$$||d_i f|| = \sqrt{\sum_{j \sim i}(df)_{ij}^2} = \sqrt{\sum_{j \sim i} w_{ij}(f_j - f_i)^2} \quad (13)$$

To avoid the zero denominators in (11), the local variation of $f$ at $i$ is defined to be

$$||d_i f|| = \sqrt{\sum_{j \sim i}(df)_{ij}^2 + \epsilon}, \quad (14)$$

where $\epsilon = 10^{-10}$.

The graph curvature is a non-linear operator.

Let $S_1(f) = \sum_i ||d_i f||$, we have the following **theorem 2**

$$D_f S_1 = \kappa f \quad (15)$$

The proof of the above theorem can be found from [27, 28]

## 6. p-Laplace operator

We define the p-Laplace operator $\Delta_p: H(V) \to H(V)$ to be

$$\Delta_p f = -\frac{1}{2} div(||df||^{p-2} df) \quad (16)$$

Clearly, $\Delta_1 = \kappa$ and $\Delta_2 = \Delta$. Next, we compute

$$(\Delta_p f)_j = \frac{1}{2} \sum_{i \sim j} \sqrt{w_{ij}} (||df||^{p-2} df_{ij} - ||df||^{p-2} df_{ji})$$
$$= \frac{1}{2} \sum_{i \sim j} \sqrt{w_{ij}} (||d_i f||^{p-2} \sqrt{w_{ij}}(f_j - f_i) - ||d_j f||^{p-2} \sqrt{w_{ij}}(f_i - f_j))$$
$$= \frac{1}{2} \sum_{i \sim j} w_{ij} (||d_i f||^{p-2} + ||d_j f||^{p-2})(f_j - f_i)$$

Thus, we have

$$(\Delta_p f)_j = \frac{1}{2}\sum_{i\sim j} w_{ij}(\|d_i f\|^{p-2} + \|d_j f\|^{p-2})(f_j - f_i) \quad (17)$$

Let $S_p(f) = \frac{1}{p}\sum_i \|d_i f\|^p$, we have the following

**theorem 3**

$$D_f S_p = p\Delta_p f \quad (18)$$

The proof of the above theorem can be found from [27, 28].

## 7. Discrete regularization on graphs

Given the network $G = (V, E, W)$. $V$ is the set of all nodes in the network and $E$ is the set of all possible interactions between these nodes. Let $y$ denote the initial ranking value in $H(V)$. $y_i$ can be defined as follows

$$y_i = \frac{1}{n}, \forall i$$

Our goal is to look for an estimated function $f$ in $H(V)$ such that $f$ is not only smooth on $G$ but also close enough to an initial function $y$. Then each node $i$ is ranked as value of $f_i$. This concept can be formulated as the following optimization problem

$$argmin_{f \in H(V)}\{S_p(f) + \frac{\mu}{2}\|f - y\|^2\} \quad (19)$$

The first term in (19) is the smoothness term. The second term is the fitting term. A positive parameter $\mu$ captures the trade-off between these two competing terms.

### 7.1 2-smoothness

When $p = 2$, the optimization problem (19) is

$$argmin_{f \in H(V)}\{\frac{1}{2}\sum_i \|d_i f\|^2 + \frac{\mu}{2}\|f - y\|^2\} \quad (20)$$

By theorem 1, we have

**Theorem 4:** The solution of (20) satisfies

$$\Delta f + \mu(f - y) = 0 \quad (21)$$

Since $\Delta$ is a linear operator, the closed form solution of (21) is

$$f = \mu(\Delta + \mu I)^{-1} y, \quad (22)$$

Where $I$ is the identity operator and $\Delta = D - W$. (22) is the algorithm proposed by [7].

### 7.2 1-smoothness

When $p = 1$, the optimization problem (19) is

$$argmin_{f \in H(V)}\{\sum_i \|d_i f\| + \frac{\mu}{2}\|f - y\|^2\}, \quad (23)$$

By theorem 2, we have

**Theorem 5:** The solution of (23) satisfies

$$\kappa f + \mu(f - y) = 0, \quad (24)$$

The curvature $\kappa$ is a non-linear operator; hence we do not have the closed form solution of equation (24). Thus, we have to construct iterative algorithm to obtain the solution. From (24), we have

$$\frac{1}{2}\sum_{i\sim j} w_{ij}\left(\frac{1}{\|d_i f\|} + \frac{1}{\|d_j f\|}\right)(f_j - f_i) + \mu(f_j - y_j) = 0 \quad (25)$$

Define the function $m: E \to R$ by

$$m_{ij} = \frac{1}{2} w_{ij}(\frac{1}{\|d_i f\|} + \frac{1}{\|d_j f\|}) \quad (26)$$

Then (25)

$$\sum_{i\sim j} m_{ij}(f_j - f_i) + \mu(f_j - y_j) = 0$$

can be transformed into

$$(\sum_{i\sim j} m_{ij} + \mu) f_j = \sum_{i\sim j} m_{ij} f_i + \mu y_j \quad (27)$$

Define the function $p: E \to R$ by

$$p_{ij} = \begin{cases} \frac{m_{ij}}{\sum_{i\sim j} m_{ij} + \mu} & \text{if } i \neq j \\ \frac{\mu}{\sum_{i\sim j} m_{ij} + \mu} & \text{if } i = j \end{cases} \quad (28)$$

Then

$$f_j = \sum_{i\sim j} p_{ij} f_i + p_{jj} y_j \quad (29)$$

Thus we can consider the iteration

$$f_j^{(t+1)} = \sum_{i\sim j} p_{ij}^{(t)} f_i^{(t)} + p_{jj}^{(t)} y_j \quad \forall j \in V$$

to obtain the solution of (23).

### 7.3 p-smoothness

For any number $p$, the optimization problem (19) is

$$argmin_{f \in H(V)}\{\frac{1}{p}\sum_i \|d_i f\|^p + \frac{\mu}{2}\|f - y\|^2\}, \quad (30)$$

By theorem 3, we have

**Theorem 6:** The solution of (30) satisfies

$$\Delta_p f + \mu(f - y) = 0, \quad (31)$$

The *p-Laplace* operator is a non-linear operator; hence we do not have the closed form solution of equation (31). Thus, we have to construct iterative algorithm to obtain the solution. From (31), we have

$$\frac{1}{2}\sum_{i\sim j} w_{ij}\left(\|d_i f\|^{p-2} + \|d_j f\|^{p-2}\right)(f_j - f_i) + \mu(f_j - y_j) = 0 \quad (32)$$

Define the function $m: E \to R$ by

$$m_{ij} = \frac{1}{2} w_{ij}(\|d_i f\|^{p-2} + \|d_j f\|^{p-2}) \quad (33)$$

Then equation (32) which is

$$\sum_{i\sim j} m_{ij}(f_j - f_i) + \mu(f_j - y_j) = 0$$

can be transformed into

$$(\sum_{i\sim j} m_{ij} + \mu) f_j = \sum_{i\sim j} m_{ij} f_i + \mu y_j \quad (34)$$

Define the function $p: E \to R$ by

$$p_{ij} = \begin{cases} \frac{m_{ij}}{\sum_{i\sim j} m_{ij} + \mu} & \text{if } i \neq j \\ \frac{\mu}{\sum_{i\sim j} m_{ij} + \mu} & \text{if } i = j \end{cases} \quad (35)$$

Then

$$f_j = \sum_{i\sim j} p_{ij} f_i + p_{jj} y_j \quad (36)$$

Thus we can consider the iteration

$$f_j^{(t+1)} = \sum_{i\sim j} p_{ij}^{(t)} f_i^{(t)} + p_{jj}^{(t)} y_j \quad \forall j \in V$$

to obtain the solution of (30).

## 8. Experiments and Results

Journal of Aviation Science & Technology                                                                                              JAST 2023## 8.1 Dataset descriptions and experimental setup

The data employed in this research was acquired from the US Bureau of Transportation Statistics (BTS) TranStats database https://www.transtats.bts.gov/DL_SelectFields.aspx?gnoyr_VQ=FGJ&QO_fu146_anzr=b0-gvzr, which is accessible publicly. The dataset that we employed in this study is from the database named "Airline On-Time Performance Data", which comprises specified records of on-time arrivals and departures for non-stop domestic flights. For this study, the original dataset contained 538,837 flights connecting 333 airports in January 2023. A graph representing the air traffic network was created to compute the network centralities such as PageRank centrality. Airports were represented as vertices of the graph, with flights between them creating the edges of the graph. The following figure shows top ten nodes with the highest ranks (using PageRank algorithm) in our constructed air traffic network:

```
[('Dallas/Fort Worth, TX', 0.03826581258848819),
 ('Denver, CO', 0.03267533509646031),
 ('Chicago, IL', 0.02829082104912407),
 ('Atlanta, GA', 0.02633732899841439),
 ('Phoenix, AZ', 0.021488552893975492),
 ('Las Vegas, NV', 0.01987526524402206),
 ('Charlotte, NC', 0.019531836046525773),
 ('Minneapolis, MN', 0.01844377010542159),
 ('Houston, TX', 0.01604298936581697),
 ('Washington, DC', 0.015721250042061331)]
```

**Figure 1:** Top ten nodes with highest PageRank scores (for air traffic network)

Our next task is to convert the real air traffic network to the line graph $G$. This line graph $G$ has 2,584 nodes and 156,121 edges.

Then, we apply the PageRank algorithm and the un-normalized graph p-Laplacian based ranking algorithm to this line graph. The top ten nodes with highest ranks obtained from the PageRank algorithm and from the un-normalized graph p-Laplacian based ranking algorithm with different $p$ values are shown in the next sub-section.

## 8.2 Experimental Results

```
[(('Dallas/Fort Worth, TX', 'Denver, CO'), 0.0007776868626699163),
 (('Chicago, IL', 'Dallas/Fort Worth, TX'), 0.000774865675043224),
 (('Chicago, IL', 'Denver, CO'), 0.0007624876094071395),
 (('Atlanta, GA', 'Dallas/Fort Worth, TX'), 0.0007609992444900775),
 (('Atlanta, GA', 'Denver, CO'), 0.0007486149361440993),
 (('Atlanta, GA', 'Chicago, IL'), 0.0007457860870315723),
 (('Dallas/Fort Worth, TX', 'Phoenix, AZ'), 0.0007353735622877435),
 (('Dallas/Fort Worth, TX', 'Las Vegas, NV'), 0.0007262451605845894),
 (('Phoenix, AZ', 'Denver, CO'), 0.0007229610274229111),
 (('Charlotte, NC', 'Dallas/Fort Worth, TX'), 0.0007221563172345392)]
```

**Figure 2:** Top ten nodes with highest PageRank scores (for the line graph)

```
[(('New York, NY', 'Nashville, TN'), 0.00038699690402477026),
 (('Milwaukee, WI', 'Dallas/Fort Worth, TX'), 0.00038699690402477016),
 (('Kona, HI', 'San Diego, CA'), 0.00038699690402477016),
 (('St. Louis, MO', 'Seattle, WA'), 0.00038699690402477016),
 (('Albany, NY', 'Miami, FL'), 0.00038699690402477701),
 (('Columbus, OH', 'Seattle, WA'), 0.00038699690402477701),
 (('Atlanta, GA', 'Detroit, MI'), 0.00038699690402477005),
 (('Myrtle Beach, SC', 'Orlando, FL'), 0.00038699690402477005),
 (('Norfolk, VA', 'Detroit, MI'), 0.00038699690402477005),
 (('St. Petersburg, FL', 'Clarksburg/Fairmont, WV'), 0.00038699690402477005)]
```

**Figure 3:** With $p = 1.1$, top ten nodes with highest ranking scores (for the line graph)

```
[(('Boise, ID', 'Burbank, CA'), 0.00038699690402472814),
 (('Washington, DC', 'Grand Rapids, MI'), 0.00038699690402472876),
 (('Dallas/Fort Worth, TX', 'Gunnison, CO'), 0.00038699690402472684),
 (('Phoenix, AZ', 'Aspen, CO'), 0.00038699690402472646),
 (('Omaha, NE', 'Seattle, WA'), 0.00038699690402472646),
 (('Nashville, TN', 'Las Vegas, NV'), 0.00038699690402472646),
 (('Minneapolis, MN', 'Des Moines, IA'), 0.00038699690402472658),
 (('Chicago, IL', 'Montrose/Delta, CO'), 0.00038699690402472658),
 (('Columbus, OH', 'Denver, CO'), 0.00038699690402472657),
 (('Chicago, IL', 'Valparaiso, FL'), 0.00038699690402472527)]
```

**Figure 4:** With $p = 1.3$, top ten nodes with highest ranking scores (for the line graph)

```
[(('Los Angeles, CA', 'Phoenix, AZ'), 0.00038699690402476533),
 (('Nashville, TN', 'Boston, MA'), 0.00038699690402476533),
 (('Fort Lauderdale, FL', 'Plattsburgh, NY'), 0.00038699690402476652),
 (('Newark, NJ', 'Bangor, ME'), 0.00038699690402476652),
 (('Sacramento, CA', 'Santa Ana, CA'), 0.00038699690402476517),
 (('Nashville, TN', 'San Antonio, TX'), 0.00038699690402476651),
 (('Orlando, FL', 'Aguadilla, PR'), 0.00038699690402476651),
 (('Los Angeles, CA', 'West Palm Beach/Palm Beach, FL'), 0.00038699690402476651),
 (('Dallas/Fort Worth, TX', 'Jackson, WY'), 0.00038699690402476651),
 (('Cincinnati, OH', 'Salt Lake City, UT'), 0.00038699690402476651)]
```

**Figure 5:** With $p = 1.5$, top ten nodes with highest ranking scores (for the line graph)

```
[(('Belleville, IL', 'St. Petersburg, FL'), 0.00038699690402485513),
 (('Atlanta, GA', 'San Juan, PR'), 0.00038699690402485125),
 (('Salt Lake City, UT', 'Oakland, CA'), 0.00038699690402485111),
 (('Richmond, VA', 'Dallas/Fort Worth, TX'), 0.00038699690402485111),
 (('Baltimore, MD', 'Syracuse, NY'), 0.00038699690402485111),
 (('San Francisco, CA', 'Redding, CA'), 0.00038699690402485111),
 (('Sanford, FL', 'Flint, MI'), 0.00038699690402485104),
 (('Sarasota/Bradenton, FL', 'Peoria, IL'), 0.00038699690402485104),
 (('Phoenix, AZ', 'San Diego, CA'), 0.00038699690402485104),
 (('Charlotte, NC', 'Omaha, NE'), 0.00038699690402485104)]
```

**Figure 6:** With $p = 1.7$, top ten nodes with highest ranking scores (for the line graph)

```
[(('Atlanta, GA', 'Shreveport, LA'), 0.00038699690402466325),
 (('Norfolk, VA', 'Jacksonville, FL'), 0.00038699690402466315),
 (('Asheville, NC', 'Dallas/Fort Worth, TX'), 0.00038699690402466266),
 (('Raleigh/Durham, NC', 'Charlotte, NC'), 0.00038699690402466266),
 (('Los Angeles, CA', 'Prescott, AZ'), 0.00038699690402466255),
 (('Detroit, MI', 'West Palm Beach/Palm Beach, FL'), 0.00038699690402466625),
 (('Baltimore, MD', 'Orlando, FL'), 0.00038699690402466244),
 (('Salt Lake City, UT', 'Colorado Springs, CO'), 0.00038699690402466244),
 (('Atlanta, GA', 'Evansville, IN'), 0.00038699690402466222),
 (('Atlanta, GA', 'Palm Springs, CA'), 0.00038699690402466222)]
```

**Figure 7:** With $p = 1.9$, top ten nodes with highest ranking scores (for the line graph)

## 8.3 Discussions

Due to lack of time and space, we cannot access the performance of the un-normalized graph p-Laplacian based ranking method. In the future, we will include not only the PageRank scores for nodes (i.e., the airports) of air traffic network dataset but also the PageRank scores and the p-Laplacian scores of edges (i.e., the flights) of the air traffic network dataset (which is nothing special but the nodes of the line graph of the air traffic network dataset) in the regression algorithm to predict the delay of flights of air traffic network dataset. But based on our previous experiences, the un-normalized graph p-Laplacian based ranking methods outperform the PageRank algorithm and its variants in a lot of classification problems and clustering problems since the p-Laplace operator of graph is the non-linear operator [8].

## 9. Conclusions

We have developed the detailed regularization frameworks for the un-normalized graph p-Laplacian ranking methods applied to rank the edges (i.e., the flights) of the air traffic network. In other words, we rank the nodes of the line graph of the air traffic network. Experiments show that the un-normalized graph p-Laplacian ranking methods are implemented successfully.

Moreover, after the ranks of the nodes and the edges are computed, we can include these two novel types of features not only in the regression algorithms (for example, linear regression algorithm) to predict the delay (in minutes) of the flights in the air traffic network but also in the classification algorithms (for example, the support vector machine algorithm) to classify whether the flights are delayed or not. This work is not included in this paper due to the lack of time and space.

Recently, to the best of my knowledge, the un-normalized hypergraph p-Laplacian based ranking methods have not yet been developed and applied to any real problems. These methods are worth researching because of their challenging nature and their close relation to partial differential equation on hypergraph field. In the scope of our future work, this hypergraph data model can represent for the air traffic network [19, 15] (Tran and Tran, 2018; Tran, 2023).